%% file: main.tex
\documentclass[twocolumn]{aastex631}
\usepackage{xcolor}
\usepackage[utf8]{inputenc}
\usepackage{amssymb,amsmath,verbatim,mathtools,needspace,enumitem,etoolbox,graphicx,physics,microtype,ulem,breqn}
\usepackage{xspace}
\usepackage[caption=false]{subfig}
\usepackage{booktabs}
\usepackage{upgreek}
\usepackage{acronym}
\usepackage{url}
\usepackage{multirow}
\usepackage{sgrmacros}

\newcommand{\sgr}{SGR~1935+2154\xspace}
\newcommand{\geo}{GEO600\xspace}
\newcommand{\xpipeline}{\texttt{X-Pipeline}\xspace}
\newcommand{\pystamp}{\texttt{PySTAMP}\xspace}

\acrodef{bbh}[BBH]{binary black hole}
\acrodef{nsbh}[NSBH]{neutron star--black hole}
\acrodef{bns}[BNS]{binary neutron star}
\acrodef{lvk}[LVK]{LIGO, Virgo, and KAGRA Collaborations}
\acrodef{chime}[CHIME]{Canadian Hydrogen Intensity Mapping Experiment}
\acrodef{chimefrb}[CHIME/FRB]{CHIME/FRB Collaboration}
\acrodef{stare2}[STARE2]{Survey for Transient Astronomical Radio Emission 2}
\acrodef{fast}[FAST]{Five-hundred-meter Aperture Spherical radio Telescope}
\acrodef{gbt}[GBT]{Green Bank Telescope}
\acrodef{evn}[EVN]{European VLBI Network}
\acrodef{em}[EM]{electromagnetic}
\acrodef{gw}[GW]{gravitational wave}
\acrodef{frb}[FRB]{fast radio burst}
\acrodef{grb}[GRB]{gamma ray burst}
\acrodef{fov}[FOV]{field-of-view}
\acrodef{snr}[S/N]{signal-to-noise ratio}
\acrodef{dm}[DM]{dispersion measure}
\acrodef{cbc}[CBC]{compact binary coalescence}
\acrodef{psd}[PSD]{power spectral density}
\acrodef{far}[FAR]{false-alarm rate}

\begin{document}

\title{A search using GEO600 for gravitational waves coincident with fast radio bursts from SGR 1935+2154}

\correspondingauthor{LIGO Scientific Collaboration, Virgo Collaboration, and KAGRA Collaboration Spokespersons}
\email{lsc-spokesperson@ligo.org, virgo-spokesperson@ego-gw.it, kscboard-chair@icrr.u-tokyo.ac.jp}

\collaboration{0}{The LIGO Scientific Collaboration, the Virgo Collaboration, and the KAGRA Collaboration}
\input{LSC-Virgo-KAGRA-Authors-Feb-2024-aas.tex}

\begin{abstract}
The magnetar SGR 1935+2154 is the only known Galactic source of fast radio bursts (FRBs).
FRBs from SGR 1935+2154 were first detected by the Canadian Hydrogen Intensity Mapping Experiment (CHIME)/FRB and the Survey for Transient Astronomical Radio Emission 2 in 2020 April, after the conclusion of the LIGO, Virgo, and KAGRA Collaborations' O3 observing run.
Here, we analyze four periods of gravitational wave (GW) data from the GEO600 detector coincident with four periods of FRB activity detected by CHIME/FRB, as well as X-ray glitches and X-ray bursts detected by NICER and NuSTAR close to the time of one of the FRBs.
We do not detect any significant GW emission from any of the events.
Instead, using a short-duration GW search (for bursts $\leq$ 1~s) we derive 50\% (90\%) upper limits of $10^{48}$ ($10^{49}$)~erg for GWs at 300~Hz and $10^{49}$ ($10^{50}$) erg at 2~kHz, and constrain the GW-to-radio energy ratio to $\leq 10^{14} - 10^{16}$.
We also derive upper limits from a long-duration search for bursts with durations between 1 and 10~s.
These represent the strictest upper limits on concurrent GW emission from FRBs.
\end{abstract}


\section{\label{sec:intro} Introduction}
\Acp{frb} are a class of extremely energetic radio transients that are theorized to be associated with neutron stars \citep{2013Sci...341...53T, 2019A&ARv..27....4P,2019PhR...821....1P, 2022Sci...378.3043B, 2023RvMP...95c5005Z}.
To date, thousands of \acp{frb} have been detected.
The majority of these have been discovered using the \ac{chime} telescope \citep{CHIME:2022dwe} by the \ac{chimefrb} \citep{2018ApJ...863...48C}\footnote{\url{https://www.chime-frb.ca/voevents}}.
Though the origins of \acp{frb} remain unknown \citep{2024Ap&SS.369...59L}, their \ac{dm} as observed by radio telescopes localizes them to extragalactic (and even cosmological) distances \citep{2007Sci...318..777L, 2017Natur.541...58C, 2019ARA&A..57..417C}.

The notable exceptions to this extragalactic consensus are the \acp{frb} associated with FRB 20200428A.
First detected in 2020 April by \ac{chimefrb} and the \ac{stare2} \citep{Bochenek:2020qry}, FRB 20200428A was quickly found to be associated with the Galactic magnetar \sgr, which was undergoing an unusual period of flaring X-ray activity at that time \citep{CHIMEFRB:2020abu, Bochenek:2020zxn, 2020GCN.27657....1B, 2020GCN.27665....1S}.
Simultaneous X-ray observations from Konus-Wind \citep{2022ATel15686....1F}, INTEGRAL \citep{Mereghetti:2020unm}, AGILE \citep{2020ATel13686....1T}, and Insight-HXMT \citep{2022ATel15698....1L} led to the first coincident observation of both radio emission and X-rays from an \ac{frb} source.
\acp{frb} from \sgr were also observed during three other epochs by \ac{chimefrb} and others, on 2020 October 08, 2022 October 14, and 2022 December 01.\footnote{
    We note that the classification of these radio bursts as \acp{frb} remains unclear: the \sgr radio bursts are a few orders of magnitude less luminous than typical extragalactic \acp{frb}, but are still brighter than most giant radio pulses \citep{2023arXiv231016932G}.
    \citet{Bochenek:2020qry} name them as \acp{frb} while \citet{2023arXiv231016932G} call them \ac{frb}-like.
    Here, we describe them as \acp{frb}.}
Additionally, X-ray glitches and bursts from \sgr were observed by NICER and NuSTAR during the nine hours surrounding the 2022 October 14 \ac{frb} \citep{Hu24}.
The connection between these X-ray bursts and \acp{frb}, even from the same magnetar, is not well understood---indeed, radio emission with no coincident X-rays has been detected from \sgr \citep{2023SciA....9F6198Z} and vice versa \citep{2017ApJ...847...85Y}.

The compact object nature of these powerful transients suggests that \acp{gw} could also be emitted by the same mechanisms that produce \acp{frb}.
The detection of \acp{gw} from an \ac{frb} source (or lack thereof) could help to elucidate the mechanisms behind \acp{frb} \citep{2023RvMP...95c5005Z}, and potentially expand the realm of detected \acp{gw} beyond those with \ac{cbc} origins.

Previous works by the \ac{lvk} have searched for \ac{gw} emission coincident with \acp{frb} \citep{LIGOScientific:2016rmm, LIGOScientific:2022jpr}, as well as for \acp{gw} from magnetar bursts \citep{LIGOScientific:2019laf, 2019ApJ...874..163A, Macquet:2021eyn, LIGOScientific:2022sts} and pulsar glitches \citep{2011PhRvD..83d2001A, 2019PhRvD.100f4058K, 2022ApJ...932..133A} using the Advanced LIGO and Advanced Virgo \ac{gw} observatories \citep{TheLIGOScientific:2014jea, TheVirgo:2014hva}.
While no detections were found in these studies, the searches have established upper limits on \ac{gw} energy that may have been emitted in association with these events.
In particular, \citet{LIGOScientific:2022jpr} performed a search for \ac{gw} emission coincident with \acp{frb} from \ac{chimefrb} during the O3a LIGO--Virgo observing run, with searches targeted at \acp{gw} from \acp{cbc} as well as generic \ac{gw} transients, setting an upper limit of $10^{51} - 10^{57}$~erg of \ac{gw} energy within 70-3560~Hz.
In addition, \citet{LIGOScientific:2022sts} placed upper limits on \ac{gw} energy ($\sim 10^{43}$ erg) coincident with 11 X-ray and soft gamma-ray magnetar bursts from \sgr.

\sgr, as the first (and at the time of writing, only) \ac{frb} source to be confidently associated with a specific neutron-star progenitor, presents a unique opportunity to search for \acp{gw} when the source is localized to a particular compact object.
Additionally, at $\sim \sgrDistance$~kpc \citep{2020ApJ...905...99Z}, it is more than two orders of magnitude nearer to Earth than the next closest \ac{frb}, which has been localized to the nearby galaxy M81, 3.6~Mpc away \citep{Bhardwaj:2021xaa, Kirsten:2021llv}.

The four periods of \ac{frb} activity from \sgr fell between the O3 and O4 observing runs of the \ac{lvk}, when the LIGO and Virgo detectors were offline\footnote{\url{https://observing.docs.ligo.org/plan/}}.
Fortunately, \geo \citep{Grote_2004, H_Lueck_2010, Affeldt_2014, Dooley_2016}, a \ac{gw} detector in Hannover, Germany that is operated by members of the LIGO Scientific Collaboration, was observing in Astrowatch mode \citep{Grote_2010} and collecting \ac{gw} data during all four periods.
The \ac{chimefrb} events for three of the four periods occurred when \geo was in observing mode, while the fourth \ac{frb} occurred within minutes of when \geo was observing (see Sec.~\ref{sec:gws}).

\begin{figure}[!htpb]
    \includegraphics[width=0.5\textwidth]{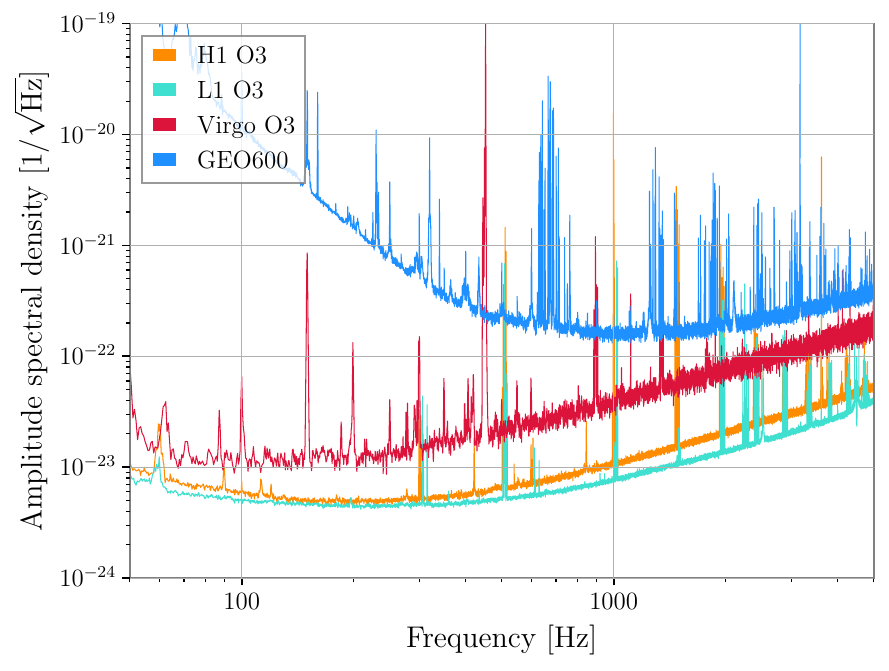}
    \caption{\label{fig:psds} Amplitude spectral density of \geo on 2020 April 28 compared to those of LIGO Hanford, LIGO Livingston, and Virgo during O3 \citep{Abbott:2020qfu}.
    While at low frequencies \geo's sensitivity is substantially diminished compared to that of the larger detectors, the gap narrows at frequencies around 2~kHz, near the expected neutron-star f-mode frequency.
    }
\end{figure}

In this paper, we analyze \geo data to search for \ac{gw} emission coincident with the four \acp{frb} observed by \ac{chimefrb} from \sgr.
We conduct two searches for unmodeled \ac{gw} transients: one targeted at short-duration bursts with $\mathcal{O}{\rm (second)}$ durations, and another aimed at long-duration bursts lasting from 1 to 10 seconds.
Due to \sgr's proximity, the results constitute the most sensitive searches for \acp{gw} from \ac{frb} sources to date, despite \geo's lower sensitivity compared to LIGO and Virgo (see Fig.~\ref{fig:psds}).
We also search for \acp{gw} coincident with the two X-ray glitches and the X-ray burst peak observed by NICER and NuSTAR in the hours around the \ac{frb} on 2022 October 14 \citep{Hu24}.
This paper is organized as follows. In Sec.~\ref{sec:frbs}, we describe the \ac{em} observations of \acp{frb} from \sgr.
Section~\ref{sec:gws} details our short- and long-duration searches for \acp{gw}, with results presented in Sec.~\ref{sec:results}.
We discuss the implications of these findings and conclude in Sec.~\ref{sec:discussion}.

\newpage
\section{\label{sec:frbs} Fast radio bursts from SGR 1935+2154}

The magnetar \sgr was discovered by Swift in 2014 \citep{2014GCN.16520....1S, 2014GCN.16522....1L}.
Since then, it has been highly active, with periods of intense emission in the X-ray and radio \citep{Israel:2016rct, 2017ApJ...847...85Y}.

On 2020 April 27, Swift observed multiple X-ray bursts from \sgr, suggesting that the magnetar had entered a period of high activity \citep{2020GCN.27657....1B}.
Less than 24 hours later, \ac{chimefrb} and \ac{stare2} detected an \ac{frb} from the location of \sgr \citep{CHIMEFRB:2020abu, Bochenek:2020zxn}.
Konus-Wind \citep{2022ATel15686....1F}, INTEGRAL \citep{Mereghetti:2020unm}, AGILE \citep{2020ATel13686....1T}, and Insight-HXMT \citep{2022ATel15698....1L} observed hard X-rays arriving at the same time, serving as the first ever observation of simultaneous radio and X-ray emission from an \ac{frb} source.
Follow-up radio observations during the same active period by the \ac{fast} \citep{2020ATel13699....1Z} and radio telescopes from the \ac{evn} \citep{Kirsten:2020yin} identified additional radio bursts from \sgr, though at lower energies.
At higher energies, no gamma-ray emission has been observed from this source \citep{2021ApJ...919..106A, 2023A&A...675A..99P}.

Since 2020 April, \sgr has had multiple periods of high activity leading to the emission of \acp{frb}.
On 2020 October 08, \ac{chimefrb} observed three \acp{frb} from \sgr arriving within a few seconds \citep{2020ATel14074....1G, 2020ATel14080....1P, 2023arXiv231016932G}.
\ac{chimefrb} and the \ac{gbt} observed \acp{frb} again two years later on 2022 October 14, with a \ac{chimefrb} event surrounded by five \ac{gbt} \acp{frb} within 1.5 seconds \citep{2022ATel15681....1D, 2022ATel15697....1M, 2023arXiv231016932G}.
During the days around the \acp{frb} on 2022 October 14, \sgr was undergoing a period of intense X-ray burst activity.
This burst storm began on October 10 \citep{Mereghetti22, Palmer2022} and was monitored by various telescopes, such as NICER, NuSTAR, and XMM-Newton (see, e.g., \citealt{Hu24, Ibrahim24}), which detected hundreds of milliseconds- to seconds-long bursts of high-energy photons.
The X-ray burst rate peaked during a flare 2.5~hr ($\pm$1~min) before the \ac{frb} and then steadily decreased over the next hours \citep{Hu24}.
In addition, the high-cadence monitoring observations allowed accurate measurements of the spin rate of \sgr (nominally 0.308~Hz; \citealt{Israel:2016rct}).
The evolution of the spin rate showed that \sgr underwent a spin-up glitch about 4.4~hr ($\pm$30~min) before the \ac{frb} and another spin-up glitch about 4.4~hr ($\pm$30~min) after the \ac{frb}, while the magnetar's spin-down rate between these two glitches was about one hundred times higher than its normal rate \citep{Hu24}.
X-ray bursts were also detected by GECAM and HEBS \citep{2022ATel15682....1W} and Konus-Wind \citep{2022ATel15686....1F} arriving within the expected \ac{frb} dispersion time.
In addition to the NICER and NuSTAR observations mentioned above \citep{2022ATel15690....1E}, Insight-HXMT \citep{2022ATel15698....1L} also observed X-rays from \sgr during this active period, though at the time of the \ac{frb} all three were occulted by the Earth.
Finally, a fourth \ac{frb} was detected by \ac{chimefrb} on 2022 December 01 \citep{2022ATel15792....1P, 2023arXiv231016932G}, accompanied by a faint hard X-ray signal detected by Fermi-GBM \citep{2022ATel15794....1Y}.

Most estimates and methods place the distance to \sgr between 1.5 and 15 kpc \citep{2013ApJ...777...14P, 2014SerAJ.189...25P, 2016ApJ...826..184S, 2018ApJ...852...54K, 2018OPhyJ...4....1R, Zhong:2020sff, 2020ApJ...905...99Z, 2021MNRAS.503.5367B}.
We adopt the determination by \citet{2020ApJ...905...99Z} of \sgrDistance~$\pm$~0.7~kpc, falling near the mean of the measurements.

\begin{figure}
    \centering
    \includegraphics[width=0.48\textwidth]{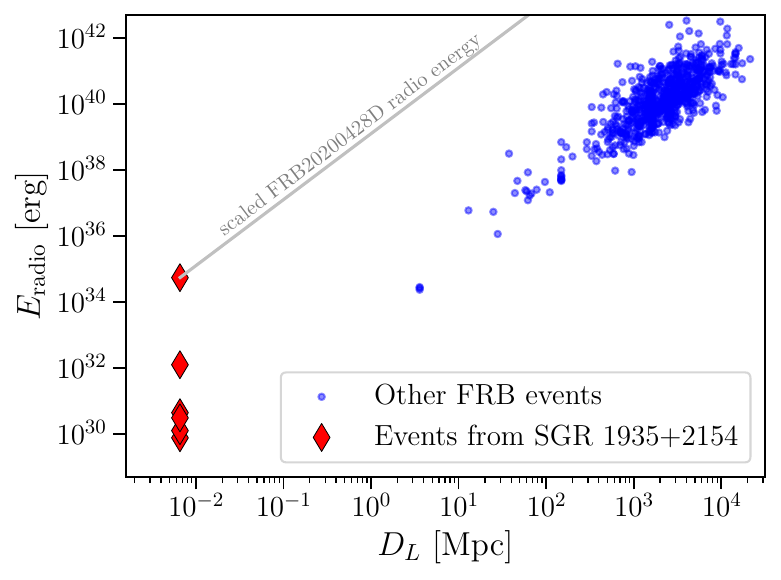}
    \caption{\label{fig:en_dist} Radio energy versus luminosity distance for the \sgr \acp{frb} investigated in this work (dark orange, \citet{2023arXiv231016932G}) and for 749 other public \acp{frb} published by \ac{chimefrb} and others \citep{2016PASA...33...45P, 2020MNRAS.495.3551R, 2021ApJS..257...59C} (blue).
    The \ac{frb} sample and the calculation of distances and radio energies is described in \citet{2023A&A...675A..99P} (with the exception of the \acp{frb} studied in \citet{LIGOScientific:2022jpr}, for which we use the lower bound 90\% distances from that analysis).
    Note that the radio energies from CHIME/FRB (derived from fluxes and fluences) should be interpreted as lower limits \citep{2021ApJS..257...59C, 2023AJ....166..138A}.
    We show the radio energy required to produce a flare as bright as that the brighest \ac{frb} from \sgr, FRB20200428D, as a function of distance.}
\end{figure}

These \sgr \acp{frb} are not quite like the rest of the population of \acp{frb}, as mentioned in Sec.~\ref{sec:intro}.
As shown in \citet{2022NatAs...6..393N}, they exhibit characteristics very similar to the extragalactic \acp{frb}, but are a few orders of magnitude less luminous.
Whether the \sgr \acp{frb} are in the tail of the same population or truly occupy a different part of the phase space remains an open question.
For example, while the 2020 April \ac{frb} from \sgr was several orders of magnitude less energetic than most \acp{frb}, it was three orders of magnitude brighter than the next brightest previously observed radio flare from a magnetar \citep{CHIMEFRB:2020abu}.
Figure~\ref{fig:en_dist} shows the radio energy as a function of distance for the \acp{frb} from \sgr, alongside a sample of 749 \acp{frb} from \ac{chimefrb}\footnote{\url{https://www.chime-frb.ca/repeaters}} \citep{2021ApJS..257...59C}, the FRBCAT \citep{2016PASA...33...45P}, and the 76~m Lovell telescope \citep{2020MNRAS.495.3551R} as collected and described in \citet{2023A&A...675A..99P}.
This could suggest that the emission mechanism which produces \acp{frb} from \sgr may be different from that which results in \acp{frb} from cosmological distances.
Despite this reduced brightness compared to the typical \ac{frb} population, \sgr's proximity as the only known Galactic \ac{frb} source means that it presents the most promising opportunity to date for multiwavelength and multimessenger studies of \ac{frb} emitters.

\subsection{\label{subsec:models} Models for coincident GW-FRB emission}
Magnetars have long been theorized to be progenitors of \acp{frb} \citep{2019PhR...821....1P}.
The detection of \acp{frb} from \sgr, a well-studied magnetar, has now confirmed this association for at least some \acp{frb}, though the exact emission mechanism remains unclear \citep{2021Univ....7...56L, 2023RvMP...95c5005Z}.

Most models which predict \ac{gw} emission from \ac{frb} progenitors assume a \ac{cbc} association, such as during or after the final stages of the \ac{cbc} inspiral \citep{2016ApJ...822L...7W, 2018PASJ...70...39Y}, long before the \ac{cbc} merger through interactions of magnetospheres \citep{2020ApJ...890L..24Z}, or other interactions of compact binaries with their environments (see \citet{2019PhR...821....1P} for a review of \ac{frb} theory).
Prior studies such as \citet{LIGOScientific:2022jpr} have searched for \acp{gw} from these sources using targeted matched-filter analyses, aimed at \ac{cbc} sources.
Since \sgr is not in a compact binary \citep{2022MNRAS.513.3550C} and has exhibited multiple periods of \ac{frb} activity, we do not expect \ac{cbc}-like \ac{gw} emission from this source.
Instead, as a magnetar, we can focus on only a few possible emission mechanisms for \acp{gw} coincident with \acp{frb}.
In particular, because \acp{gw} are induced by time-varying quadrupole moments, we review models which predict \ac{em} magnetar activity associated with such moments.

Since at least some \acp{frb} originate from magnetars, theories have drawn connections between them and magnetar giant flares \citep{2016ApJ...827...59T, 2020ApJ...899L..27M, 2024MNRAS.528.5323C}, which are are thought to be powered by magnetic activity near the surface of a neutron star \citep{1996ApJ...473..322T, 2005Natur.434.1104G}.
These giant flares are rare but so energetic that \ac{gw} emission may be detectable due to hydromagnetic coupling of the magnetic dipole to the mass quadrupole \citep{2001MNRAS.327..639I, 2011PhRvD..83j4014C}.
Quasi-periodic oscillations in the X-ray tails of giant flares may also create \acp{gw} through torsional or Alfv\'en modes which alter the star's quadrupole moment \citep{2011MNRAS.418..659L, 2014MNRAS.439.1522G, 2017CQGra..34p4002Q}.
While no giant flares from \sgr were detected during its periods of \ac{frb} activity, the coincident X-ray activity suggests a potential link in the provenance of the high-energy \ac{em} emission.

Crustal f-modes are a possible source of transient \acp{gw} from isolated neutron stars \citep{2018ASSL..457..673G, 2020PhRvD.101j3009H}.
These typically fall at around 2~kHz \citep{1996PhRvL..77.4134A}, near the frequencies where \geo is most sensitive (see Fig.~\ref{fig:psds}).

Moreover, neutron-star glitches, such as those from \sgr in 2022 October investigated in this work, may emit \acp{gw} potentially observable by current \ac{gw} detectors \citep{2011PhRvD..84b3007P, 2012MNRAS.423.2058W, 2015ApJ...807..132M}.
Previous limits were derived for the Vela pulsar (located at 290~pc; \citet{2003ApJ...596.1137D}) glitch in 2006, providing limits on the emitted \ac{gw} energy of the order of 10$^{45}$ erg \citep{2011PhRvD..83d2001A}.

\section{Search for gravitational waves}\label{sec:gws}
Using data from \geo, we search for generic \ac{gw} transients from \sgr around the times of four \acp{frb} detected by \ac{chimefrb}.
\geo is a dual-recycled Michelson interferometer with folded arms and takes astrophysical observations in the 40 Hz - 6 kHz frequency band when operating in Astrowatch mode \citep{Grote_2010}.
Over the past two decades, it has pioneered several key technologies for \ac{gw} detectors \citep{Affeldt_2014, PhysRevLett.126.041102}.
\geo has lower sensitivity compared to the Advanced~LIGO and Advanced~Virgo detectors ($\sim 10^{-22}/\sqrt{\mathrm{Hz}}$ at 1 kHz, see Fig.~\ref{fig:psds}), but has strengths in uptime.
It continued taking observations during the initial period of the COVID-19 pandemic and subsequent LIGO and Virgo upgrades throughout 2020-2022, during which \ac{chimefrb} observed these \acp{frb} from \sgr.

Given the unknown nature of the emission mechanism of the \acp{frb}, we search for generic transient gravitational-wave signals present in the \geo data using two unmodeled burst searches: \pystamp \citep{2021PhRvD.104j2005M}, targeted at long-duration bursts with lengths from 1 to 10~s, and \xpipeline \citep{Sutton:2009gi, Was:2012zq}, for short-duration bursts lasting less than 1~s.
Previous searches for \acp{gw} coincident with \acp{frb}, such as the ones presented in \citet{LIGOScientific:2022jpr}, also considered a possible \ac{cbc} origin for the \ac{gw} emission; the non-compact binary nature of \sgr precludes the use of \ac{cbc} matched-filter searches.
Given that \sgr is a magnetar, we follow the previous \ac{gw} magnetar study presented in \citet{LIGOScientific:2022sts} and employ \pystamp to perform a long-duration search, which has not previously been used for \ac{gw} \ac{frb} analyses.
Additionally, prior searches have typically been restricted to coincidences with \acp{frb} where data from at least two \ac{gw} detectors is available.
For these \acp{frb} from \sgr, only \geo was observing, so we employ a single-detector search.
This limits our ability to veto candidates based on coherence between detectors and the amount of background that can be estimated, reducing the search sensitivity, but given the extraordinary nature of these \acp{frb}, we determined that a single-detector search in \geo data was warranted.

\begin{table*}
    \centering
    \caption{\label{tab:events} Table of \ac{frb} and X-ray events for which we perform \ac{gw} searches.
    The long-duration \pystamp search is performed with one time window, while the short-duration \xpipeline search is performed with both a compact window and an extended window, where data availability and timing uncertainties permit.
    A dash (-) indicates that no search was performed, while an asterisk (*) denotes search windows which were necessarily truncated due to data availability.}
    \begin{tabular}{@{}lllllll@{}}
    \toprule
                                &          & Window for                  & Compact window for        & Extended window for \\
    FRB/X-ray event             & Time (UTC) & long-duration search (s)  & short-duration search (s) & short-duration search (s)\\
    \midrule
    \midrule
    2020 April 28                  & 14:34:24 & $[-1200, 120]$ & $[-4, 4]$ & $[-1200, 120]$ \\
    2020 October 08                  & 02:23:33 & $[-1200, 120]$ & $[-4, 4]$ & $[-1200, 120]$ \\
    2022 October 14 X-ray glitch 1   & 15:07:12 & -            & -       & $[-480, -240]$* \\
    2022 October 14 X-ray burst peak & 16:55:12 & -            & -       & $[-1000, 1000]$ \\
    2022 October 14                  & 19:21:39 & $[-687, 120]$* & $[-4, 4]$ & $[-600, 120]$* \\
    2022 October 14 X-ray glitch 2   & 23:45:36 & -            & -       & $[-500, 500]$*  \\
    2022 December 01                 & 22:06:59 & -            & -       & $[-1200, -400]$* \\
    \bottomrule
    \end{tabular}
\end{table*}

For the 2020 April 28, 2020 October 08, and 2022 October 14 \acp{frb}, we perform a search within an ``on-source'' time window starting at 1200~s before the infinite-frequency arrival time (i.e., the time accounting for the frequency-dependent delay introduced by the dispersion measure) of the \ac{frb}, $t_0$, and ending 120~s after.
This asymmetric window is motivated by the expectation that any potential \acp{gw} are likely generated from the interior of the magnetar, preceding \ac{frb} emission from the magnetosphere or beyond.
On 2020 October 08, three bursts were detected by \ac{chimefrb} within 3~s; we use the first time, corresponding to the \ac{frb} with the highest fluence on that day, as our $t_0$.
The \ac{frb} on 2022 December 01 occurred during a time when \geo was not taking data, having exited observing mode approximately six minutes before the \ac{frb}, at 22:01:09 UTC.
\geo returned to observing mode 23 minutes later, at 22:23:59 UTC.
To be consistent with the on-source window for the other \acp{frb}, we analyze the 800~s period of data beginning 1200~s before the \ac{frb} and ending shortly before \geo exited observing mode.
Due to the large uncertainties in \ac{frb}--\ac{gw} models (as described above in Sec.~\ref{subsec:models}), we use a wide extended on-source window of $[-1200, 120]$~s.
This allows us to probe a broad parameter space while keeping the detector behavior relatively stationary.
We also employ a compact $[-4, 4]$~s search window in the short-duration search to probe the time immediately surrounding the \ac{frb} with higher sensitivity.
In addition, we search for emission around the time of three X-ray events detected by NuSTAR and NICER on 2022 October 14.
These events consisted of a spin-up glitch, a peak in the X-ray burst emission, and another spin-up glitch.
Since the uncertainty in their times is greater than a minute, we only perform an extended-window search, targeting a symmetric $[-1000, 1000]$~s window, subject to data availability.
Table~\ref{tab:events} summarizes the times and windows for which we perform searches.

We restrict our search to frequencies from 300~Hz to 4096~Hz, with the lower cutoff set by the low frequency sensitivity of \geo and the upper cutoff aiming to capture neutron-star crustal f-modes, which are predicted to fall at approximately 2~kHz \citep{1996PhRvL..77.4134A}.

\subsection{\label{subsec:injs} Simulated waveforms to quantify sensitivity}
We measure the sensitivity of our search by inserting simulated waveforms (``injections'') into the off-source data and quantifying the pipeline's ability to recover them.
For the short-duration \xpipeline search, these injections are largely the same as those used in \citet{LIGOScientific:2022jpr}.
The waveforms include Sine-Gaussians and damped sinusoids and are summarized in Table~\ref{tab:injections}.
We briefly describe each waveform family below:
\begin{itemize}
    \item \textbf{Sine-Gaussians:} The majority of the simulated waveforms we use are Sine-Gaussians, which can model starquakes and certain neutron-star f-modes.
    They are described in Eq.~1 of \citet{2017ApJ...841...89A}.
    Most of these injections are performed with inclinations chosen randomly, but we also employ some optimally inclined (circular polarization only, emitted face-on to the observer) waveforms near the expected f-modes at $\sim$2000~Hz to better constrain our sensitivity to these models.
    In all the injected waveforms, we use a quality factor $Q = 9$ (the approximate number of cycles in the waveform) following \citet{LIGOScientific:2020lst, LIGOScientific:2022jpr}, with central frequencies $f_0$ spanning from 300~Hz to 3560~Hz, as shown in Table~\ref{tab:injections}.

    \item \textbf{Damped sinusoids:} We also use damped sinusoids to characterize any ringdown behavior in the magnetar.
    The waveform is described in Eq.~C12 of \citet{LIGOScientific:2022sts}.
    These are placed at two frequencies, 1590~Hz and 2020~Hz, to represent plausible f-mode signals.
    For each frequency, we use two damping timescales to probe a larger parameter space.
\end{itemize}
The waveforms used by the long-duration \pystamp analysis are also Sine-Gaussians, but with a duration parameter of 10~s.
They are also described in Table~\ref{tab:injections}.

\begin{table}[!htbp]
  \caption{\label{tab:injections} Parameters for waveforms injected into off-source data for recovery to quantify each search's sensitivity.
  For the generic short-duration transient search \xpipeline, we follow the labeling convention in \citet{LIGOScientific:2022jpr} for each waveform, where ``SG'' waveforms are sine-Gaussians and ``DS2P'' (damped sinusoid 2 polarizations) waveforms represent ringdowns.
  There are few enough long-duration waveforms that we did not assign labels to them.
  The duration parameter scales the width of the Gaussian envelope for the sine-Gaussian chirplets, and describes the damping time of the damped sinusoids used as ringdown waveforms.
  The $^c$ superscript denotes waveforms with circular polarizations.
  }
    \begin{tabular}{lcc}
\toprule
 Label &   Frequency $f_0$          & Duration Parameter \\
      &    [Hz]             &    [ms]      \\
      \midrule
      \midrule
      \multicolumn{3}{c}{Short-duration Sine--Gaussian Chirplets}\\
      \midrule
SG-D &         300             &  3.3       \\
SG-E &         500             &  0.20       \\
SG-F &         1100            &  0.91       \\
SG-G &         1600            &  0.63       \\
SG-H &         1995            &  0.50       \\
SG-I &         2600            &  0.38       \\
SG-J &         3100            &  0.32       \\
SG-K &         3560            &  0.28       \\
SG-L$^c$ &         1600            &  0.63       \\
SG-M$^c$ &         1995            &  0.50       \\
      \midrule
            \multicolumn{3}{c}{Short-duration Ringdowns}\\
                  \midrule
DS2P-A&         1590            &  100      \\
DS2P-B&         1590            &  200      \\
DS2P-C&         2020            &  100      \\
DS2P-D&         2020            &  200      \\
        \midrule
\multicolumn{3}{c}{Long-duration Sine--Gaussian Chirplets}\\
\midrule
- &         520             &  $10^4$      \\
- &         1020             & $10^4$       \\
- &         1520            &  $10^4$       \\
- &         2020            &  $10^4$       \\
      \bottomrule
\end{tabular}
\end{table}

\subsection{\label{subsec:pystamp} Long-duration search with \pystamp}
We use \pystamp \citep{2021PhRvD.104j2005M} to target \ac{gw} signals with durations longer than $1$~s around the three \acp{frb} with coincident \geo data.
The background distribution and the detection efficiency of the search are characterized using an off-source window that consists of $\sim 12$~hr of data centered on the event time, excluding the on-source window described above.
The workflow of the pipeline is as follows.
The data are first down-sampled from $16384$~Hz to $8192$~Hz, and then high-pass filtered with a frequency cutoff of $40$~Hz to remove potential spectral leakage from lower frequencies.
After these preprocessing steps, the resulting time series are split into $1$~s Hann-windowed segments with $50 \%$ overlap.
The fast Fourier transform is computed over each segment to build a time-frequency map (tf-map) with a resolution of $1$~s $\times 1$~Hz.
For each frequency bin, the \ac{psd} is estimated by taking the median of the squared modulus of the Fourier transform over $1320$~s of adjacent data (similar to Welch's method but using the median instead of the mean), and a \ac{snr} tf-map is built by dividing the value of the Fourier transform in each pixel by the square root of the \ac{psd}.
To identify candidate \ac{gw} events, a pattern recognition algorithm is run over the tf-map.
We use the \texttt{burstegard} algorithm \citep{Pr2016} which identifies clusters of neighboring pixels whose \ac{snr} is above a threshold of $2.5$.
Clusters consisting of $5$ or more pixels are saved as candidate \ac{gw} events.
Each cluster is then assessed a ranking statistic $\Lambda$ that is the sum of the \ac{snr} of each of its pixels divided by the square root of the total number of pixels.

Clusters found in the off-source window form the background of the search and are used to estimate the \ac{far} of clusters found in the on-source window as a function of the ranking statistic $\Lambda$.
\pystamp is primarily intended to work on cross-correlated data from a pair of independent detectors, which allows for the simulation of an extended amount of background by shifting the time series of one detector with respect to the other.
Such a method cannot be applied here in the single-detector \geo search.
Hence, the background lifetime is limited to the duration of the off-source window, so the \ac{far} of each cluster can only be estimated down to a minimum of $\sim1$ per $12$~hours ($2.3 \times 10^{-5}$~Hz).
See Sec.~\ref{sec:discussion} for further discussion of the limited \ac{far}.

Noise from \ac{gw} detectors typically features narrowband spectral artifacts that appear as horizontal lines in a time-frequency representation (or vertical lines in the amplitude spectral density; see Fig.~\ref{fig:psds}).
Because the \ac{psd} is estimated for each frequency bin by taking the median over neighboring time segments, most of these lines are correctly factored into the \ac{psd} and do not generate high \ac{snr} pixels.
However, we observe an excess of clusters in the off-source window around some specific frequencies, likely due to fluctuations of spectral lines around their central values.
We therefore remove clusters for a narrow range of frequencies corresponding to known \geo spectral lines.
In order to reject short, broadband noise transients (known as glitches), we also remove clusters for which more than $30\%$ of the total energy is contained within a single $1$~s time segment.

\subsection{\label{subsec:xpipeline} Short-duration search with \xpipeline}
We perform a search for short-duration unmodeled \ac{gw} transients using \xpipeline \citep{Sutton:2009gi, Was:2012zq}.
While typically run as a coherent search across multiple detectors (such as in previous searches for \acp{gw} from \acp{frb} \citep{LIGOScientific:2022jpr}, \acp{grb} \citep{LIGOScientific:2020lst, LIGOScientific:2021iyk}, and magnetars \citep{LIGOScientific:2022sts}), we use \xpipeline in a single-detector mode because \geo was the only \ac{gw} detector collecting data at the time of the \acp{frb}.
\xpipeline splits the \ac{psd}-whitened data into 64~s segments, then applies a Fourier transform to produce time-frequency maps.
The time-frequency pixels with amplitudes in the highest 1\% that neighbor each other are clustered into candidate detection events.
Each event is then assigned a ranking statistic based on the summed energy contained in the pixels.
To determine the significance of these candidate events, we compare them against a distribution of background energies empirically measured in an identical manner from an ``off-source'' period.
This is chosen to fall around (but not including) the time of the on-source data and to be long enough to allow for meaningful significances to be calculated but not so long that the detector's behavior is nonstationary.
We employ a 24-hour off-source window, symmetric about each event's time.

When \xpipeline is run on data from multiple detectors as is typical, vetoes of problematic event candidates can be applied by utilizing the presumed coherence of any real \ac{gw} event across detectors.
This is unfortunately not an option in a single-detector search such as this, meaning that the search becomes more vulnerable to background noise.
To improve the sensitivity of our search, we apply frequency-domain vetoes based on the distribution of time-frequency candidate events in the off-source window for each \ac{frb} search.
We veto narrow frequency bands ($\sim 10$~Hz bandwidth) where there is considerable excess noise; most of these vetoes corresponded to known spectral lines from the \geo detector.

\section{\label{sec:results} Search results and limits on coincident emission}
In this section, we present the results of the long- and short-duration searches described above (see Table~\ref{tab:events} for a summary).

\begin{figure*}[!htbp]
    \centering
    \includegraphics[width=0.95\textwidth]{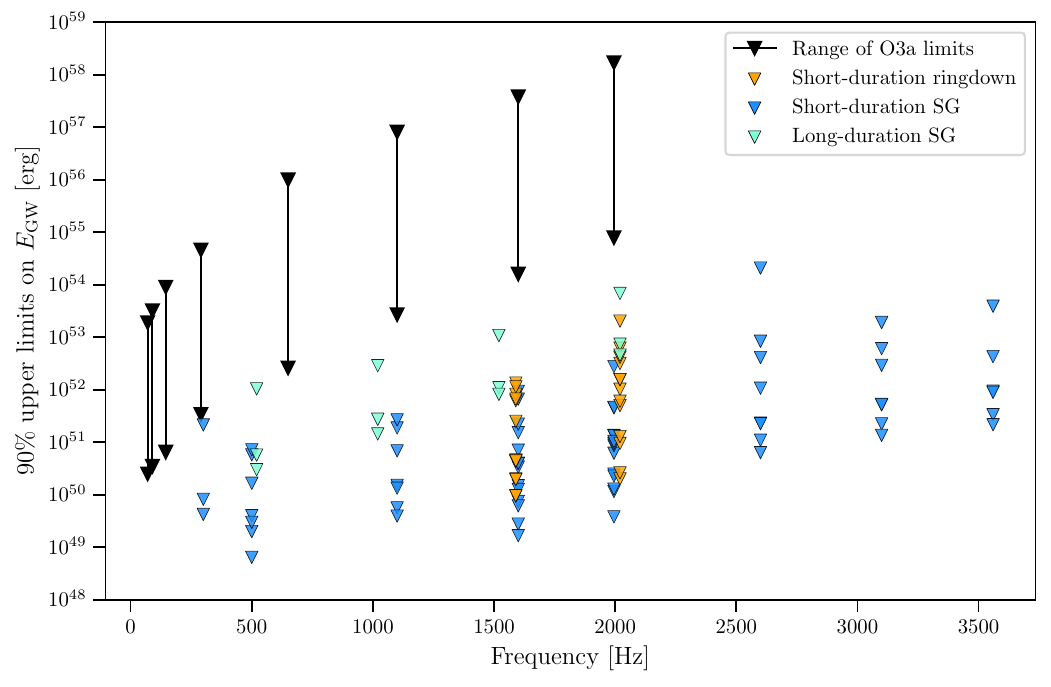}
    \caption{\label{fig:egw_compared} The 90\% upper limits on the emitted \ac{gw} energy from \sgr coincident with \acp{frb}, alongside \ac{gw} energy limits from \acp{frb} during the O3a observing run as reported in \citet{LIGOScientific:2022jpr}.
    We plot the short-duration ringdown (orange) and Sine-Gaussian (SG; blue) waveforms from Table~\ref{tab:limits-short-90}, and the long-duration SG waveforms (aquamarine) from Table~\ref{tab:limits-long}.
    The previous range of 90\% limits from \citet{LIGOScientific:2022jpr}, based on the lower bounds of the 90\% credible distance as reported in their Table~A1, are plotted in vertical black lines.
    These are for a short-duration search; \citet{LIGOScientific:2022jpr} did not perform a long-duration analysis.
    }
\end{figure*}

\input{longdurTable.tex}
\input{shortdurTable}

We do not find any candidate \ac{gw} events in either the long-duration or short-duration compact-window searches for any of the \acp{frb}.
For the long-duration \pystamp search, no triggers survive the cuts described in Sec.~\ref{subsec:pystamp} for the 2020 April and 2022 October \acp{frb}.
Five triggers survive for the 2020 October \ac{frb}, but the loudest trigger has a \ac{far} of $\sim 1$ per 1000~s with a p-value of 0.76 and is thus not significant.
In the short-duration \xpipeline case, only the 2020 April, 2020 October, and 2022 October \acp{frb} compact-window searches had enough background to be considered useful for \ac{gw} searches, as described above.
The only surviving trigger from these three searches is from the 2022 October \ac{frb} and has a p-value of 0.53, and thus is also not significant.

For the short-duration extended-window searches, long off-source windows are required to estimate the background, and the single-detector nature of this search prevents the use of ``time-slides'' between multiple detectors to artificially generate background.
These limitations mean that for the extended-window short-duration searches, there are sometimes as few as six off-source background trials, limiting any potential detection's maximum significance to a p-value of $1/6 \approx 0.17$---insufficient for any meaningful statement about the astrophysical nature of any outlier in the data.
Thus, instead of reporting potential \ac{gw} candidates from the extended-window searches with inconsequential statements about significance, we decided to use the searches only to determine the loudest trigger within the on-source window for a given waveform model, thereby setting an upper limit on the corresponding \ac{gw} energy.
This is the ``loudest event statistic'' \citep{2009CQGra..26q5009B, 2013CQGra..30g9502B}.

For all searches and windows (listed in Table~\ref{tab:events}), we estimate the root-sum-square signal amplitude of the \ac{gw} strain $h_{\rm rss}$ (see Eqs.~3 and 4 of \citet{2013arXiv1304.0210S}) at 50\% and 90\% detection efficiency to set upper limits on the \ac{gw} energy emitted.
To convert the $h_{\rm rss}$ values for each injection type that are output by the search pipelines to energy, we use (following \citet{2013arXiv1304.0210S})
\begin{equation}
    E_{\rm GW} = \frac{2}{5}\frac{\pi^2 c^3}{G} D_L^2 f_0^2 h_{\rm rss}^2,
    \label{eq:energy}
\end{equation}
with $f_0$ describing the central frequency of each injection and $D_L$ set to the \sgrDistance~kpc distance of \sgr \citep{2020ApJ...905...99Z}.
The results of the long-duration search are shown in Tab.~\ref{tab:limits-long}.
The 50\% and 90\% limits from the short-duration analysis are shown in Tab.~\ref{tab:limits-short-50} and Tab.~\ref{tab:limits-short-90} respectively.
We show in Fig.~\ref{fig:egw_compared} our 90\% upper limits from both the long- and short-duration analyses as a function of frequency, corresponding to the upper limits at which 90\% of the injected signals were recovered.
To be explicit, the $X$\% $h_{\rm rss}$ value for a given waveform model (and corresponding \ac{gw} energy) is calculated by finding the $h_{\rm rss}$ at which $X$\% of the injected waveforms are recovered (i.e., found with a significance higher than the loudest non-injection event in the window).

For some injections in the short-duration search (mostly those at lower frequencies such as SG-D with $f_0 = 300$~Hz), limits could not be established because noise in the detector prevented sufficient recovery of the injected signals.
Limited data availability and poor data quality around the time of the 2020 October 08 event meant that no injection reached 90\% recovery, leading to the lack of 90\% limits.

\begin{figure*}[!htbp]
    \centering
    \includegraphics[width=0.95\textwidth]{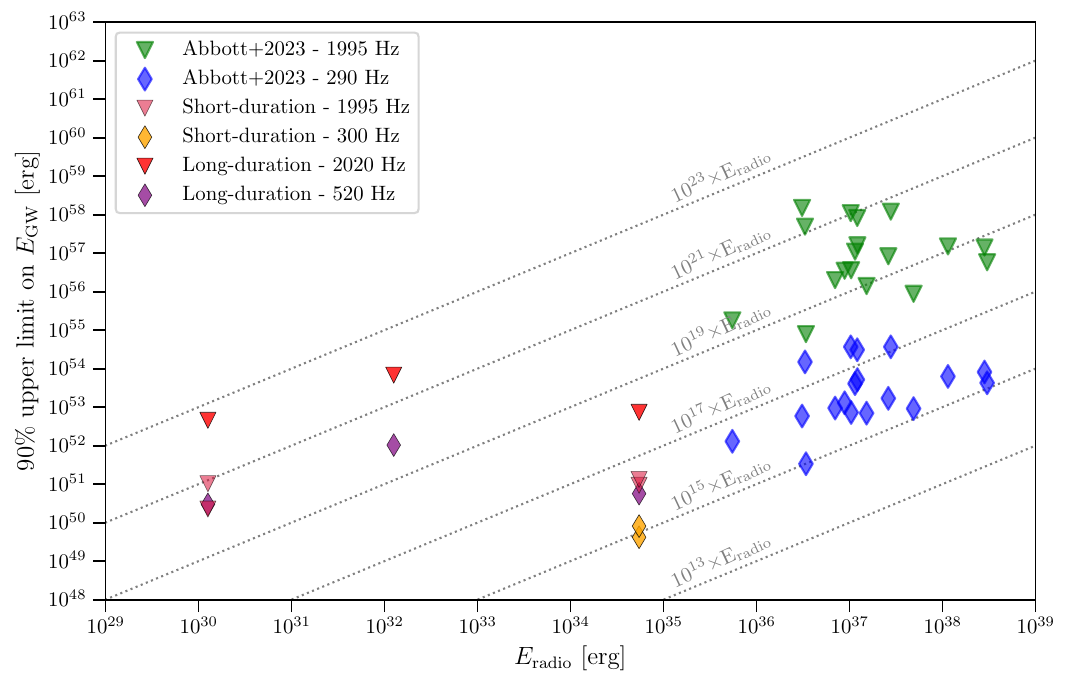}
    \caption{\label{fig:egw_eradio} 90\% upper limits on the emitted \ac{gw} energy from \acp{frb} as a function of the \ac{frb}'s radio energy.
    In the pink, orange, red, and purple, we show limits from \acp{frb} emitted by \sgr, for both our short- and long-duration searches.
    We plot limits for the Sine-Gaussian model at 300~Hz (SG-D) and 1995~Hz (SG-H) for the short-duration search and at 520~Hz and 2020~Hz for the long-duration search.
    In the blue and green markers, we show the upper limits on \ac{gw} energy and the corresponding radio energy for \acp{frb} analyzed in \citet{LIGOScientific:2022jpr}, at 290~Hz and 1995~Hz, for events with radio flux/fluence information from \citet{2021ApJS..257...59C} allowing for radio energy reconstruction.
    The estimated radio energies are calculated as described in \citet{2023A&A...675A..99P}, scaled to the lower bound 90\% distances as reported in \citet{LIGOScientific:2022jpr}.
    Note that the radio energies (derived from fluxes and fluences) should be interpreted as lower limits \citep{2021ApJS..257...59C, 2023AJ....166..138A}.
    We also plot dotted lines representing different ratios of $E_{\rm GW}$ to $E_{\rm radio}$, showing a slight improvement in $E_{\rm GW} / E_{\rm radio}$ compared to the \citet{LIGOScientific:2022jpr} results.}
\end{figure*}

The nondetection of \ac{gw} emission from our analyzed \acp{frb} implies that the \ac{gw}-to-radio energy ratio must be less than $E_{\rm GW} / E_{\rm radio} \sim8\times10^{14}$ at the 90\% level, for a time window from [-4, 4]~s at a \ac{gw} frequency of approximately 300~Hz.
At the $\sim 2$~kHz frequencies close to the neutron-star f-mode, $E_{\rm GW} / E_{\rm radio} \lesssim 1.7 \times 10^{16}$ at the 90\% level.
The \ac{gw} and radio energies for our analyzed \acp{frb} are shown in Fig.~\ref{fig:egw_eradio}, alongside the same quantities for \acp{frb} from O3a analyzed in \citet{LIGOScientific:2022jpr}.

\section{\label{sec:discussion} Discussion and conclusion}
The previous best limits on coincident \ac{gw} emission with \acp{frb} were set by \citet{LIGOScientific:2022jpr} using extragalactic \acp{frb} observed during the O3a \ac{lvk} observing run, using SG waveforms and \xpipeline.
Fig.~\ref{fig:egw_compared} shows the 90\% upper limits on the \ac{gw} energy during our analyzed \acp{frb} from \sgr, compared previous limits presented as a range spanning the best and worst 90\% limits from \citet{LIGOScientific:2022jpr}.
In the short-duration search at approximately 300~Hz, the best previous 90\% upper limit on \ac{gw} energy was set at $3.4 \times 10^{51}$~erg; at approximately 2~kHz, the best previous 90\% upper limit was $7.9 \times 10^{54}$~erg.
We improve on the 300~Hz constraint by about two orders of magnitude, and the 2~kHz constraint by over four orders of magnitude.
No previous long-duration searches around \ac{frb} events have been performed, so the long-duration search results presented here represent the first constraints on such emission.
Studies have predicted that magnetar flares can emit up to $10^{48}-10^{49}$~erg in \ac{gw} energy near the f-mode for $\sim 200$~ms \citep{2001MNRAS.327..639I, 2011PhRvD..83j4014C}---a regime that is probed by our most stringent 50\% short-duration constraints at approximately 2~kHz (see SG-H waveform in Table~\ref{tab:limits-short-50}).
We also slightly improve the upper limit on $E_{\rm GW} / E_{\rm radio}$, as shown in Fig.~\ref{fig:egw_eradio}.

We note that our results are not the most constraining limits on \ac{gw} emission from the magnetar \sgr, which are reported in \citet{LIGOScientific:2022sts} in a search for \ac{gw} emission around times of magnetar X-ray and gamma-ray flares.
The relationship between these magnetar bursts and \acp{frb} is poorly understood, but are likely to be caused by different physical processes, even if the underlying magnetar behavior may be related \citep{2024MNRAS.530.1885T}.
Hence, both \ac{gw} limits are complementary and can help to better understand the emission mechanisms at play.
Considering the X-ray spin-up glitches, our best limits on the emitted \ac{gw} energy ($\sim10^{51}$ erg at 300~Hz) are still far from the X-ray measured changes in rotational energy ($\sim10^{42}$ erg from \citet{Hu24}).

Since $E_{\rm GW} \propto D_L^2$ in Eq.~\ref{eq:energy}, our constraints are heavily dependent on the distance to \sgr.
As mentioned in Sec.~\ref*{sec:frbs}, estimates for \sgr's distance vary by almost an order of magnitude.
If its true distance is as close as the 1.5~kpc measured by \citet{2021MNRAS.503.5367B}, our energy constraints would improve by a factor of almost 20.
On the other hand, if \sgr is at almost 15~kpc, as suggested by \citet{2016ApJ...826..184S}, our constraints would worsen by a factor of 5.

As a single-detector search, the number of possible background trials (and thus the assessment of \ac{gw} candidates via p-value or \ac{far}) is limited by the time in which the detector behavior remains similar to that of the on-source window.
For example, the minimum \ac{far} for the long-duration \pystamp search was limited to $\sim$ 1 per 12 hours.
Since we do not recover any candidates in the on-source windows which appear to differentiate themselves from the background, we did not have to assess any candidate's significance beyond 1 per 12 hours, meaning that this limitation did not affect our results.
However, future single-detector searches may encounter the problem where a candidate in the on-source window is unlike anything found in the background trials, complicating an accurate assessment of its significance.
Some \ac{gw} \ac{cbc} searches have implemented techniques to improve single-detector significance estimation \citep{2019arXiv190108580S, 2022CQGra..39u5012C}; unmodeled searches like \pystamp and \xpipeline may also benefit from such enhancements.

At the time of writing, no \acp{frb} have been detected from \sgr since 2022.
The O4 observing run of the \ac{lvk}, with participation from the LIGO, Virgo, and KAGRA detectors will continue until mid-2025.
Given the increased sensitivity of these detectors compared to \geo, any \sgr \ac{frb} during the remainder of O4 could provide another opportunity to probe the \ac{gw}--\ac{frb} connection.

\vspace{7pt}
\input{P2000488_v24_edited.tex}

\facilities{\geo, CHIME, NICER, NuSTAR}

\software{
    \texttt{astropy} \citep{astropy:2013},
    \texttt{gwpy} \citep{gwpy-software},
    \texttt{LVK Algorithm Library Suite} \citep{lalsuite}.
    \texttt{matplotlib} \citep{Hunter:2007},
    \texttt{numpy} \citep{harris2020array},
    \texttt{pandas} \citep{reback2020pandas, mckinney2010data}
}
\bibliography{coinc_frb}{}
\bibliographystyle{aasjournal}

\allauthors

\end{document}

%% file: LSC-Virgo-KAGRA-Authors-Feb-2024-aas.tex
\author[0000-0003-4786-2698]{A.~G.~Abac}
\affiliation{Max Planck Institute for Gravitational Physics (Albert Einstein Institute), D-14476 Potsdam, Germany}
\author{R.~Abbott}
\affiliation{LIGO Laboratory, California Institute of Technology, Pasadena, CA 91125, USA}
\author{I.~Abouelfettouh}
\affiliation{LIGO Hanford Observatory, Richland, WA 99352, USA}
\author{F.~Acernese}
\affiliation{Dipartimento di Farmacia, Universit\`a di Salerno, I-84084 Fisciano, Salerno, Italy}
\affiliation{INFN, Sezione di Napoli, I-80126 Napoli, Italy}
\author[0000-0002-8648-0767]{K.~Ackley}
\affiliation{University of Warwick, Coventry CV4 7AL, United Kingdom}
\author{S.~Adhicary}
\affiliation{The Pennsylvania State University, University Park, PA 16802, USA}
\author[0000-0002-4559-8427]{N.~Adhikari}
\affiliation{University of Wisconsin-Milwaukee, Milwaukee, WI 53201, USA}
\author[0000-0002-5731-5076]{R.~X.~Adhikari}
\affiliation{LIGO Laboratory, California Institute of Technology, Pasadena, CA 91125, USA}
\author{V.~K.~Adkins}
\affiliation{Louisiana State University, Baton Rouge, LA 70803, USA}
\author[0000-0002-8735-5554]{D.~Agarwal}
\affiliation{Universit\'e catholique de Louvain, B-1348 Louvain-la-Neuve, Belgium}
\affiliation{Inter-University Centre for Astronomy and Astrophysics, Pune 411007, India}
% ...

%% file: longdurTable.tex
\begin{table*}
    \centering
    \footnotesize
    \caption{\label{tab:limits-long} The 50\% and 90\% upper limits on \ac{gw} emission energy in ergs from the long-duration \pystamp search.
        Each frequency corresponds to the central frequency $f_0$ of a Sine-Gaussian waveform with duration parameter equal to 10~s.}
    \begin{tabular}{lllllllll}
    \toprule
    Event                 & \multicolumn{2}{c}{520 Hz} & \multicolumn{2}{c}{1020 Hz} & \multicolumn{2}{c}{1520 Hz} & \multicolumn{2}{c}{2020 Hz} \\
                          & 50\% & 90\% & 50\% & 90\% & 50\% & 90\% & 50\% & 90\% \\
    \midrule
    \midrule
    2020 April 28 & \LongDurTwentyTwentyAprilFRBSineGaussianFiveTwentyHzULFifty  & \LongDurTwentyTwentyAprilFRBSineGaussianFiveTwentyHzULNinety  & \LongDurTwentyTwentyAprilFRBSineGaussianTenTwentyHzULFifty  & \LongDurTwentyTwentyAprilFRBSineGaussianTenTwentyHzULNinety  & \LongDurTwentyTwentyAprilFRBSineGaussianFifteenTwentyHzULFifty  & \LongDurTwentyTwentyAprilFRBSineGaussianFifteenTwentyHzULNinety  & \LongDurTwentyTwentyAprilFRBSineGaussianTwentyTwentyHzULFifty  & \LongDurTwentyTwentyAprilFRBSineGaussianTwentyTwentyHzULNinety  \\
    2020 October 08 & \LongDurTwentyTwentyOctFRBSineGaussianFiveTwentyHzULFifty    & \LongDurTwentyTwentyOctFRBSineGaussianFiveTwentyHzULNinety    & \LongDurTwentyTwentyOctFRBSineGaussianTenTwentyHzULFifty    & \LongDurTwentyTwentyOctFRBSineGaussianTenTwentyHzULNinety    & \LongDurTwentyTwentyOctFRBSineGaussianFifteenTwentyHzULFifty    & \LongDurTwentyTwentyOctFRBSineGaussianFifteenTwentyHzULNinety    & \LongDurTwentyTwentyOctFRBSineGaussianTwentyTwentyHzULFifty    & \LongDurTwentyTwentyOctFRBSineGaussianTwentyTwentyHzULNinety    \\
    2022 October 14 & \LongDurTwentyTwentyTwoOctFRBSineGaussianFiveTwentyHzULFifty & \LongDurTwentyTwentyTwoOctFRBSineGaussianFiveTwentyHzULNinety & \LongDurTwentyTwentyTwoOctFRBSineGaussianTenTwentyHzULFifty & \LongDurTwentyTwentyTwoOctFRBSineGaussianTenTwentyHzULNinety & \LongDurTwentyTwentyTwoOctFRBSineGaussianFifteenTwentyHzULFifty & \LongDurTwentyTwentyTwoOctFRBSineGaussianFifteenTwentyHzULNinety & \LongDurTwentyTwentyTwoOctFRBSineGaussianTwentyTwentyHzULFifty & \LongDurTwentyTwentyTwoOctFRBSineGaussianTwentyTwentyHzULNinety \\
    \bottomrule
    \end{tabular}
\end{table*}

%% file: shortdurTable.tex
\begin{table*}
    \centering
    \footnotesize
    \caption{\label{tab:limits-short-50} The 50\% upper limits on \ac{gw} emission energy in ergs from the short-duration \xpipeline search.
    Each injection waveform is defined as in Table~\ref{tab:injections}.
    Dashes indicate that no limit was obtainable for this set of injections due to insufficient background or poor data quality.}
    \begin{tabular}{lllllllll}
    \toprule
    Event             &    & SG-D & SG-E & SG-F & SG-G & SG-H & SG-I & SG-J \\
    Date        & Window   & 300 Hz & 500 Hz & 1100 Hz & 1600 Hz & 1995 Hz & 2600 Hz & 3100 Hz \\
    \midrule
    \midrule
    2020 April 28              & Compact      & \ShortDurTwentyTwentyAprilFRBShortSineGaussianThreeHundredHzULFifty   & \ShortDurTwentyTwentyAprilFRBShortSineGaussianFiveHundredHzULFifty   & \ShortDurTwentyTwentyAprilFRBShortSineGaussianElevenHundredHzULFifty   & \ShortDurTwentyTwentyAprilFRBShortSineGaussianSixteenHundredHzULFifty   & \ShortDurTwentyTwentyAprilFRBShortSineGaussianNineteenNinetyFiveHzULFifty   & \ShortDurTwentyTwentyAprilFRBShortSineGaussianTwentySixHundredHzULFifty   & \ShortDurTwentyTwentyAprilFRBShortSineGaussianThirtyOneHundredHzULFifty   \\
    2020 April 28              & Extended   & \ShortDurTwentyTwentyAprilFRBLongSineGaussianThreeHundredHzULFifty    & \ShortDurTwentyTwentyAprilFRBLongSineGaussianFiveHundredHzULFifty    & \ShortDurTwentyTwentyAprilFRBLongSineGaussianElevenHundredHzULFifty    & \ShortDurTwentyTwentyAprilFRBLongSineGaussianSixteenHundredHzULFifty    & \ShortDurTwentyTwentyAprilFRBLongSineGaussianNineteenNinetyFiveHzULFifty    & \ShortDurTwentyTwentyAprilFRBLongSineGaussianTwentySixHundredHzULFifty    & \ShortDurTwentyTwentyAprilFRBLongSineGaussianThirtyOneHundredHzULFifty    \\
    2020 October 08             & Compact      & \ShortDurTwentyTwentyOctFRBShortSineGaussianThreeHundredHzULFifty     & \ShortDurTwentyTwentyOctFRBShortSineGaussianFiveHundredHzULFifty     & \ShortDurTwentyTwentyOctFRBShortSineGaussianElevenHundredHzULFifty     & \ShortDurTwentyTwentyOctFRBShortSineGaussianSixteenHundredHzULFifty     & \ShortDurTwentyTwentyOctFRBShortSineGaussianNineteenNinetyFiveHzULFifty     & \ShortDurTwentyTwentyOctFRBShortSineGaussianTwentySixHundredHzULFifty     & \ShortDurTwentyTwentyOctFRBShortSineGaussianThirtyOneHundredHzULFifty     \\
    2020 October 08             & Extended   & \ShortDurTwentyTwentyOctFRBLongSineGaussianThreeHundredHzULFifty      & \ShortDurTwentyTwentyOctFRBLongSineGaussianFiveHundredHzULFifty      & \ShortDurTwentyTwentyOctFRBLongSineGaussianElevenHundredHzULFifty      & \ShortDurTwentyTwentyOctFRBLongSineGaussianSixteenHundredHzULFifty      & \ShortDurTwentyTwentyOctFRBLongSineGaussianNineteenNinetyFiveHzULFifty      & \ShortDurTwentyTwentyOctFRBLongSineGaussianTwentySixHundredHzULFifty      & \ShortDurTwentyTwentyOctFRBLongSineGaussianThirtyOneHundredHzULFifty      \\
    2022 October 14 glitch 1   & Extended  & \ShortDurTwentyTwentyTwoOctGlitchOneSineGaussianThreeHundredHzULFifty & \ShortDurTwentyTwentyTwoOctGlitchOneSineGaussianFiveHundredHzULFifty & \ShortDurTwentyTwentyTwoOctGlitchOneSineGaussianElevenHundredHzULFifty & \ShortDurTwentyTwentyTwoOctGlitchOneSineGaussianSixteenHundredHzULFifty & \ShortDurTwentyTwentyTwoOctGlitchOneSineGaussianNineteenNinetyFiveHzULFifty & \ShortDurTwentyTwentyTwoOctGlitchOneSineGaussianTwentySixHundredHzULFifty & \ShortDurTwentyTwentyTwoOctGlitchOneSineGaussianThirtyOneHundredHzULFifty \\
    2022 October 14 X-ray peak & Extended & \ShortDurTwentyTwentyTwoOctXrayPeakSineGaussianThreeHundredHzULFifty  & \ShortDurTwentyTwentyTwoOctXrayPeakSineGaussianFiveHundredHzULFifty  & \ShortDurTwentyTwentyTwoOctXrayPeakSineGaussianElevenHundredHzULFifty  & \ShortDurTwentyTwentyTwoOctXrayPeakSineGaussianSixteenHundredHzULFifty  & \ShortDurTwentyTwentyTwoOctXrayPeakSineGaussianNineteenNinetyFiveHzULFifty  & \ShortDurTwentyTwentyTwoOctXrayPeakSineGaussianTwentySixHundredHzULFifty  & \ShortDurTwentyTwentyTwoOctXrayPeakSineGaussianThirtyOneHundredHzULFifty  \\
    2022 October 14            & Compact      & \ShortDurTwentyTwentyTwoOctFRBShortSineGaussianThreeHundredHzULFifty  & \ShortDurTwentyTwentyTwoOctFRBShortSineGaussianFiveHundredHzULFifty  & \ShortDurTwentyTwentyTwoOctFRBShortSineGaussianElevenHundredHzULFifty  & \ShortDurTwentyTwentyTwoOctFRBShortSineGaussianSixteenHundredHzULFifty  & \ShortDurTwentyTwentyTwoOctFRBShortSineGaussianNineteenNinetyFiveHzULFifty  & \ShortDurTwentyTwentyTwoOctFRBShortSineGaussianTwentySixHundredHzULFifty  & \ShortDurTwentyTwentyTwoOctFRBShortSineGaussianThirtyOneHundredHzULFifty  \\
    2022 October 14            & Extended   & \ShortDurTwentyTwentyTwoOctFRBLongSineGaussianThreeHundredHzULFifty   & \ShortDurTwentyTwentyTwoOctFRBLongSineGaussianFiveHundredHzULFifty   & \ShortDurTwentyTwentyTwoOctFRBLongSineGaussianElevenHundredHzULFifty   & \ShortDurTwentyTwentyTwoOctFRBLongSineGaussianSixteenHundredHzULFifty   & \ShortDurTwentyTwentyTwoOctFRBLongSineGaussianNineteenNinetyFiveHzULFifty   & \ShortDurTwentyTwentyTwoOctFRBLongSineGaussianTwentySixHundredHzULFifty   & \ShortDurTwentyTwentyTwoOctFRBLongSineGaussianThirtyOneHundredHzULFifty   \\
    2022 October 14 glitch 2   & Extended & \ShortDurTwentyTwentyTwoOctGlitchTwoSineGaussianThreeHundredHzULFifty & \ShortDurTwentyTwentyTwoOctGlitchTwoSineGaussianFiveHundredHzULFifty & \ShortDurTwentyTwentyTwoOctGlitchTwoSineGaussianElevenHundredHzULFifty & \ShortDurTwentyTwentyTwoOctGlitchTwoSineGaussianSixteenHundredHzULFifty & \ShortDurTwentyTwentyTwoOctGlitchTwoSineGaussianNineteenNinetyFiveHzULFifty & \ShortDurTwentyTwentyTwoOctGlitchTwoSineGaussianTwentySixHundredHzULFifty & \ShortDurTwentyTwentyTwoOctGlitchTwoSineGaussianThirtyOneHundredHzULFifty \\
    2022 December 1            & Extended   & \ShortDurTwentyTwentyTwoDecFRBLongSineGaussianThreeHundredHzULFifty   & \ShortDurTwentyTwentyTwoDecFRBLongSineGaussianFiveHundredHzULFifty   & \ShortDurTwentyTwentyTwoDecFRBLongSineGaussianElevenHundredHzULFifty   & \ShortDurTwentyTwentyTwoDecFRBLongSineGaussianSixteenHundredHzULFifty   & \ShortDurTwentyTwentyTwoDecFRBLongSineGaussianNineteenNinetyFiveHzULFifty   & \ShortDurTwentyTwentyTwoDecFRBLongSineGaussianTwentySixHundredHzULFifty   & \ShortDurTwentyTwentyTwoDecFRBLongSineGaussianThirtyOneHundredHzULFifty   \\
    \bottomrule
    Event            &     & SG-K & SG-L & SG-M & DS2P-A & DS2P-B & DS2P-C & DS2P-D \\
    Date        & Window   & 3560 Hz & 1600 Hz & 1995 Hz & 1590 Hz & 1590 Hz & 2020 Hz & 2020 Hz \\
    \midrule
    \midrule
    2020 April 28              & Compact      & \ShortDurTwentyTwentyAprilFRBShortSineGaussianThirtyFiveSixtyHzULFifty   & \ShortDurTwentyTwentyAprilFRBShortSineGaussianSixteenHundredCircHzULFifty   & \ShortDurTwentyTwentyAprilFRBShortSineGaussianNineteenNinetyFiveCircHzULFifty   & \ShortDurTwentyTwentyAprilFRBShortRingdownFifteenNinetyaHzULFifty   & \ShortDurTwentyTwentyAprilFRBShortRingdownFifteenNinetybHzULFifty   & \ShortDurTwentyTwentyAprilFRBShortRingdownTwentyTwentyaHzULFifty   & \ShortDurTwentyTwentyAprilFRBShortRingdownTwentyTwentybHzULFifty   \\
    2020 April 28              & Extended   & \ShortDurTwentyTwentyAprilFRBLongSineGaussianThirtyFiveSixtyHzULFifty    & \ShortDurTwentyTwentyAprilFRBLongSineGaussianSixteenHundredCircHzULFifty    & \ShortDurTwentyTwentyAprilFRBLongSineGaussianNineteenNinetyFiveCircHzULFifty    & \ShortDurTwentyTwentyAprilFRBLongRingdownFifteenNinetyaHzULFifty    & \ShortDurTwentyTwentyAprilFRBLongRingdownFifteenNinetybHzULFifty    & \ShortDurTwentyTwentyAprilFRBLongRingdownTwentyTwentyaHzULFifty    & \ShortDurTwentyTwentyAprilFRBLongRingdownTwentyTwentybHzULFifty    \\
    2020 October 08             & Compact      & \ShortDurTwentyTwentyOctFRBShortSineGaussianThirtyFiveSixtyHzULFifty     & \ShortDurTwentyTwentyOctFRBShortSineGaussianSixteenHundredCircHzULFifty     & \ShortDurTwentyTwentyOctFRBShortSineGaussianNineteenNinetyFiveCircHzULFifty     & \ShortDurTwentyTwentyOctFRBShortRingdownFifteenNinetyaHzULFifty     & \ShortDurTwentyTwentyOctFRBShortRingdownFifteenNinetybHzULFifty     & \ShortDurTwentyTwentyOctFRBShortRingdownTwentyTwentyaHzULFifty     & \ShortDurTwentyTwentyOctFRBShortRingdownTwentyTwentybHzULFifty     \\
    2020 October 08             & Extended   & \ShortDurTwentyTwentyOctFRBLongSineGaussianThirtyFiveSixtyHzULFifty      & \ShortDurTwentyTwentyOctFRBLongSineGaussianSixteenHundredCircHzULFifty      & \ShortDurTwentyTwentyOctFRBLongSineGaussianNineteenNinetyFiveCircHzULFifty      & \ShortDurTwentyTwentyOctFRBLongRingdownFifteenNinetyaHzULFifty      & \ShortDurTwentyTwentyOctFRBLongRingdownFifteenNinetybHzULFifty      & \ShortDurTwentyTwentyOctFRBLongRingdownTwentyTwentyaHzULFifty      & \ShortDurTwentyTwentyOctFRBLongRingdownTwentyTwentybHzULFifty      \\
    2022 October 14 glitch 1   & Extended  & \ShortDurTwentyTwentyTwoOctGlitchOneSineGaussianThirtyFiveSixtyHzULFifty & \ShortDurTwentyTwentyTwoOctGlitchOneSineGaussianSixteenHundredCircHzULFifty & \ShortDurTwentyTwentyTwoOctGlitchOneSineGaussianNineteenNinetyFiveCircHzULFifty & \ShortDurTwentyTwentyTwoOctGlitchOneRingdownFifteenNinetyaHzULFifty & \ShortDurTwentyTwentyTwoOctGlitchOneRingdownFifteenNinetybHzULFifty & \ShortDurTwentyTwentyTwoOctGlitchOneRingdownTwentyTwentyaHzULFifty & \ShortDurTwentyTwentyTwoOctGlitchOneRingdownTwentyTwentybHzULFifty \\
    2022 October 14 X-ray peak & Extended & \ShortDurTwentyTwentyTwoOctXrayPeakSineGaussianThirtyFiveSixtyHzULFifty  & \ShortDurTwentyTwentyTwoOctXrayPeakSineGaussianSixteenHundredCircHzULFifty  & \ShortDurTwentyTwentyTwoOctXrayPeakSineGaussianNineteenNinetyFiveCircHzULFifty  & \ShortDurTwentyTwentyTwoOctXrayPeakRingdownFifteenNinetyaHzULFifty  & \ShortDurTwentyTwentyTwoOctXrayPeakRingdownFifteenNinetybHzULFifty  & \ShortDurTwentyTwentyTwoOctXrayPeakRingdownTwentyTwentyaHzULFifty  & \ShortDurTwentyTwentyTwoOctXrayPeakRingdownTwentyTwentybHzULFifty  \\
    2022 October 14            & Compact      & \ShortDurTwentyTwentyTwoOctFRBShortSineGaussianThirtyFiveSixtyHzULFifty  & \ShortDurTwentyTwentyTwoOctFRBShortSineGaussianSixteenHundredCircHzULFifty  & \ShortDurTwentyTwentyTwoOctFRBShortSineGaussianNineteenNinetyFiveCircHzULFifty  & \ShortDurTwentyTwentyTwoOctFRBShortRingdownFifteenNinetyaHzULFifty  & \ShortDurTwentyTwentyTwoOctFRBShortRingdownFifteenNinetybHzULFifty  & \ShortDurTwentyTwentyTwoOctFRBShortRingdownTwentyTwentyaHzULFifty  & \ShortDurTwentyTwentyTwoOctFRBShortRingdownTwentyTwentybHzULFifty  \\
    2022 October 14            & Extended   & \ShortDurTwentyTwentyTwoOctFRBLongSineGaussianThirtyFiveSixtyHzULFifty   & \ShortDurTwentyTwentyTwoOctFRBLongSineGaussianSixteenHundredCircHzULFifty   & \ShortDurTwentyTwentyTwoOctFRBLongSineGaussianNineteenNinetyFiveCircHzULFifty   & \ShortDurTwentyTwentyTwoOctFRBLongRingdownFifteenNinetyaHzULFifty   & \ShortDurTwentyTwentyTwoOctFRBLongRingdownFifteenNinetybHzULFifty   & \ShortDurTwentyTwentyTwoOctFRBLongRingdownTwentyTwentyaHzULFifty   & \ShortDurTwentyTwentyTwoOctFRBLongRingdownTwentyTwentybHzULFifty   \\
    2022 October 14 glitch 2   & Extended & \ShortDurTwentyTwentyTwoOctGlitchTwoSineGaussianThirtyFiveSixtyHzULFifty & \ShortDurTwentyTwentyTwoOctGlitchTwoSineGaussianSixteenHundredCircHzULFifty & \ShortDurTwentyTwentyTwoOctGlitchTwoSineGaussianNineteenNinetyFiveCircHzULFifty & \ShortDurTwentyTwentyTwoOctGlitchTwoRingdownFifteenNinetyaHzULFifty & \ShortDurTwentyTwentyTwoOctGlitchTwoRingdownFifteenNinetybHzULFifty & \ShortDurTwentyTwentyTwoOctGlitchTwoRingdownTwentyTwentyaHzULFifty & \ShortDurTwentyTwentyTwoOctGlitchTwoRingdownTwentyTwentybHzULFifty \\
    2022 December 1            & Extended   & \ShortDurTwentyTwentyTwoDecFRBLongSineGaussianThirtyFiveSixtyHzULFifty   & \ShortDurTwentyTwentyTwoDecFRBLongSineGaussianSixteenHundredCircHzULFifty   & \ShortDurTwentyTwentyTwoDecFRBLongSineGaussianNineteenNinetyFiveCircHzULFifty   & \ShortDurTwentyTwentyTwoDecFRBLongRingdownFifteenNinetyaHzULFifty   & \ShortDurTwentyTwentyTwoDecFRBLongRingdownFifteenNinetybHzULFifty   & \ShortDurTwentyTwentyTwoDecFRBLongRingdownTwentyTwentyaHzULFifty   & \ShortDurTwentyTwentyTwoDecFRBLongRingdownTwentyTwentybHzULFifty   \\
    \bottomrule
    \end{tabular}
\end{table*}

\begin{table*}
    \centering
    \footnotesize
    \caption{\label{tab:limits-short-90} Same as Table~\ref{tab:limits-short-50} but with 90\% upper limits.
    For some injections (marked by a dash), fewer than 90\% of the injections are recovered, preventing the calculation of 90\% limits.}
    \begin{tabular}{lllllllll}
    \toprule
    Event           &      & SG-D & SG-E & SG-F & SG-G & SG-H & SG-I & SG-J \\
    Date        & Window   & 300 Hz & 500 Hz & 1100 Hz & 1600 Hz & 1995 Hz & 2600 Hz & 3100 Hz \\
    \midrule
    \midrule
    2020 April 28              & Compact      & \ShortDurTwentyTwentyAprilFRBShortSineGaussianThreeHundredHzULNinety   & \ShortDurTwentyTwentyAprilFRBShortSineGaussianFiveHundredHzULNinety   & \ShortDurTwentyTwentyAprilFRBShortSineGaussianElevenHundredHzULNinety   & \ShortDurTwentyTwentyAprilFRBShortSineGaussianSixteenHundredHzULNinety   & \ShortDurTwentyTwentyAprilFRBShortSineGaussianNineteenNinetyFiveHzULNinety   & \ShortDurTwentyTwentyAprilFRBShortSineGaussianTwentySixHundredHzULNinety   & \ShortDurTwentyTwentyAprilFRBShortSineGaussianThirtyOneHundredHzULNinety   \\
    2020 April 28              & Extended   & \ShortDurTwentyTwentyAprilFRBLongSineGaussianThreeHundredHzULNinety    & \ShortDurTwentyTwentyAprilFRBLongSineGaussianFiveHundredHzULNinety    & \ShortDurTwentyTwentyAprilFRBLongSineGaussianElevenHundredHzULNinety    & \ShortDurTwentyTwentyAprilFRBLongSineGaussianSixteenHundredHzULNinety    & \ShortDurTwentyTwentyAprilFRBLongSineGaussianNineteenNinetyFiveHzULNinety    & \ShortDurTwentyTwentyAprilFRBLongSineGaussianTwentySixHundredHzULNinety    & \ShortDurTwentyTwentyAprilFRBLongSineGaussianThirtyOneHundredHzULNinety    \\
    2020 October 08             & Compact      & \ShortDurTwentyTwentyOctFRBShortSineGaussianThreeHundredHzULNinety     & \ShortDurTwentyTwentyOctFRBShortSineGaussianFiveHundredHzULNinety     & \ShortDurTwentyTwentyOctFRBShortSineGaussianElevenHundredHzULNinety     & \ShortDurTwentyTwentyOctFRBShortSineGaussianSixteenHundredHzULNinety     & \ShortDurTwentyTwentyOctFRBShortSineGaussianNineteenNinetyFiveHzULNinety     & \ShortDurTwentyTwentyOctFRBShortSineGaussianTwentySixHundredHzULNinety     & \ShortDurTwentyTwentyOctFRBShortSineGaussianThirtyOneHundredHzULNinety     \\
    2020 October 08             & Extended   & \ShortDurTwentyTwentyOctFRBLongSineGaussianThreeHundredHzULNinety      & \ShortDurTwentyTwentyOctFRBLongSineGaussianFiveHundredHzULNinety      & \ShortDurTwentyTwentyOctFRBLongSineGaussianElevenHundredHzULNinety      & \ShortDurTwentyTwentyOctFRBLongSineGaussianSixteenHundredHzULNinety      & \ShortDurTwentyTwentyOctFRBLongSineGaussianNineteenNinetyFiveHzULNinety      & \ShortDurTwentyTwentyOctFRBLongSineGaussianTwentySixHundredHzULNinety      & \ShortDurTwentyTwentyOctFRBLongSineGaussianThirtyOneHundredHzULNinety      \\
    2022 October 14 glitch 1   & Extended  & \ShortDurTwentyTwentyTwoOctGlitchOneSineGaussianThreeHundredHzULNinety & \ShortDurTwentyTwentyTwoOctGlitchOneSineGaussianFiveHundredHzULNinety & \ShortDurTwentyTwentyTwoOctGlitchOneSineGaussianElevenHundredHzULNinety & \ShortDurTwentyTwentyTwoOctGlitchOneSineGaussianSixteenHundredHzULNinety & \ShortDurTwentyTwentyTwoOctGlitchOneSineGaussianNineteenNinetyFiveHzULNinety & \ShortDurTwentyTwentyTwoOctGlitchOneSineGaussianTwentySixHundredHzULNinety & \ShortDurTwentyTwentyTwoOctGlitchOneSineGaussianThirtyOneHundredHzULNinety \\
    2022 October 14 X-ray peak & Extended & \ShortDurTwentyTwentyTwoOctXrayPeakSineGaussianThreeHundredHzULNinety  & \ShortDurTwentyTwentyTwoOctXrayPeakSineGaussianFiveHundredHzULNinety  & \ShortDurTwentyTwentyTwoOctXrayPeakSineGaussianElevenHundredHzULNinety  & \ShortDurTwentyTwentyTwoOctXrayPeakSineGaussianSixteenHundredHzULNinety  & \ShortDurTwentyTwentyTwoOctXrayPeakSineGaussianNineteenNinetyFiveHzULNinety  & \ShortDurTwentyTwentyTwoOctXrayPeakSineGaussianTwentySixHundredHzULNinety  & \ShortDurTwentyTwentyTwoOctXrayPeakSineGaussianThirtyOneHundredHzULNinety  \\
    2022 October 14            & Compact      & \ShortDurTwentyTwentyTwoOctFRBShortSineGaussianThreeHundredHzULNinety  & \ShortDurTwentyTwentyTwoOctFRBShortSineGaussianFiveHundredHzULNinety  & \ShortDurTwentyTwentyTwoOctFRBShortSineGaussianElevenHundredHzULNinety  & \ShortDurTwentyTwentyTwoOctFRBShortSineGaussianSixteenHundredHzULNinety  & \ShortDurTwentyTwentyTwoOctFRBShortSineGaussianNineteenNinetyFiveHzULNinety  & \ShortDurTwentyTwentyTwoOctFRBShortSineGaussianTwentySixHundredHzULNinety  & \ShortDurTwentyTwentyTwoOctFRBShortSineGaussianThirtyOneHundredHzULNinety  \\
    2022 October 14            & Extended   & \ShortDurTwentyTwentyTwoOctFRBLongSineGaussianThreeHundredHzULNinety   & \ShortDurTwentyTwentyTwoOctFRBLongSineGaussianFiveHundredHzULNinety   & \ShortDurTwentyTwentyTwoOctFRBLongSineGaussianElevenHundredHzULNinety   & \ShortDurTwentyTwentyTwoOctFRBLongSineGaussianSixteenHundredHzULNinety   & \ShortDurTwentyTwentyTwoOctFRBLongSineGaussianNineteenNinetyFiveHzULNinety   & \ShortDurTwentyTwentyTwoOctFRBLongSineGaussianTwentySixHundredHzULNinety   & \ShortDurTwentyTwentyTwoOctFRBLongSineGaussianThirtyOneHundredHzULNinety   \\
    2022 October 14 glitch 2   & Extended & \ShortDurTwentyTwentyTwoOctGlitchTwoSineGaussianThreeHundredHzULNinety & \ShortDurTwentyTwentyTwoOctGlitchTwoSineGaussianFiveHundredHzULNinety & \ShortDurTwentyTwentyTwoOctGlitchTwoSineGaussianElevenHundredHzULNinety & \ShortDurTwentyTwentyTwoOctGlitchTwoSineGaussianSixteenHundredHzULNinety & \ShortDurTwentyTwentyTwoOctGlitchTwoSineGaussianNineteenNinetyFiveHzULNinety & \ShortDurTwentyTwentyTwoOctGlitchTwoSineGaussianTwentySixHundredHzULNinety & \ShortDurTwentyTwentyTwoOctGlitchTwoSineGaussianThirtyOneHundredHzULNinety \\
    2022 December 1            & Extended   & \ShortDurTwentyTwentyTwoDecFRBLongSineGaussianThreeHundredHzULNinety   & \ShortDurTwentyTwentyTwoDecFRBLongSineGaussianFiveHundredHzULNinety   & \ShortDurTwentyTwentyTwoDecFRBLongSineGaussianElevenHundredHzULNinety   & \ShortDurTwentyTwentyTwoDecFRBLongSineGaussianSixteenHundredHzULNinety   & \ShortDurTwentyTwentyTwoDecFRBLongSineGaussianNineteenNinetyFiveHzULNinety   & \ShortDurTwentyTwentyTwoDecFRBLongSineGaussianTwentySixHundredHzULNinety   & \ShortDurTwentyTwentyTwoDecFRBLongSineGaussianThirtyOneHundredHzULNinety   \\
    \bottomrule
    Event          &       & SG-K & SG-L & SG-M & DS2P-A & DS2P-B & DS2P-C & DS2P-D \\
    Date        & Window   & 3560 Hz & 1600 Hz & 1995 Hz & 1590 Hz & 1590 Hz & 2020 Hz & 2020 Hz \\
    \midrule
    \midrule
    2020 April 28              & Compact      & \ShortDurTwentyTwentyAprilFRBShortSineGaussianThirtyFiveSixtyHzULNinety   & \ShortDurTwentyTwentyAprilFRBShortSineGaussianSixteenHundredCircHzULNinety   & \ShortDurTwentyTwentyAprilFRBShortSineGaussianNineteenNinetyFiveCircHzULNinety   & \ShortDurTwentyTwentyAprilFRBShortRingdownFifteenNinetyaHzULNinety   & \ShortDurTwentyTwentyAprilFRBShortRingdownFifteenNinetybHzULNinety   & \ShortDurTwentyTwentyAprilFRBShortRingdownTwentyTwentyaHzULNinety   & \ShortDurTwentyTwentyAprilFRBShortRingdownTwentyTwentybHzULNinety   \\
    2020 April 28              & Extended   & \ShortDurTwentyTwentyAprilFRBLongSineGaussianThirtyFiveSixtyHzULNinety    & \ShortDurTwentyTwentyAprilFRBLongSineGaussianSixteenHundredCircHzULNinety    & \ShortDurTwentyTwentyAprilFRBLongSineGaussianNineteenNinetyFiveCircHzULNinety    & \ShortDurTwentyTwentyAprilFRBLongRingdownFifteenNinetyaHzULNinety    & \ShortDurTwentyTwentyAprilFRBLongRingdownFifteenNinetybHzULNinety    & \ShortDurTwentyTwentyAprilFRBLongRingdownTwentyTwentyaHzULNinety    & \ShortDurTwentyTwentyAprilFRBLongRingdownTwentyTwentybHzULNinety    \\
    2020 October 08             & Compact      & \ShortDurTwentyTwentyOctFRBShortSineGaussianThirtyFiveSixtyHzULNinety     & \ShortDurTwentyTwentyOctFRBShortSineGaussianSixteenHundredCircHzULNinety     & \ShortDurTwentyTwentyOctFRBShortSineGaussianNineteenNinetyFiveCircHzULNinety     & \ShortDurTwentyTwentyOctFRBShortRingdownFifteenNinetyaHzULNinety     & \ShortDurTwentyTwentyOctFRBShortRingdownFifteenNinetybHzULNinety     & \ShortDurTwentyTwentyOctFRBShortRingdownTwentyTwentyaHzULNinety     & \ShortDurTwentyTwentyOctFRBShortRingdownTwentyTwentybHzULNinety     \\
    2020 October 08             & Extended   & \ShortDurTwentyTwentyOctFRBLongSineGaussianThirtyFiveSixtyHzULNinety      & \ShortDurTwentyTwentyOctFRBLongSineGaussianSixteenHundredCircHzULNinety      & \ShortDurTwentyTwentyOctFRBLongSineGaussianNineteenNinetyFiveCircHzULNinety      & \ShortDurTwentyTwentyOctFRBLongRingdownFifteenNinetyaHzULNinety      & \ShortDurTwentyTwentyOctFRBLongRingdownFifteenNinetybHzULNinety      & \ShortDurTwentyTwentyOctFRBLongRingdownTwentyTwentyaHzULNinety      & \ShortDurTwentyTwentyOctFRBLongRingdownTwentyTwentybHzULNinety      \\
    2022 October 14 glitch 1   & Extended  & \ShortDurTwentyTwentyTwoOctGlitchOneSineGaussianThirtyFiveSixtyHzULNinety & \ShortDurTwentyTwentyTwoOctGlitchOneSineGaussianSixteenHundredCircHzULNinety & \ShortDurTwentyTwentyTwoOctGlitchOneSineGaussianNineteenNinetyFiveCircHzULNinety & \ShortDurTwentyTwentyTwoOctGlitchOneRingdownFifteenNinetyaHzULNinety & \ShortDurTwentyTwentyTwoOctGlitchOneRingdownFifteenNinetybHzULNinety & \ShortDurTwentyTwentyTwoOctGlitchOneRingdownTwentyTwentyaHzULNinety & \ShortDurTwentyTwentyTwoOctGlitchOneRingdownTwentyTwentybHzULNinety \\
    2022 October 14 X-ray peak & Extended & \ShortDurTwentyTwentyTwoOctXrayPeakSineGaussianThirtyFiveSixtyHzULNinety  & \ShortDurTwentyTwentyTwoOctXrayPeakSineGaussianSixteenHundredCircHzULNinety  & \ShortDurTwentyTwentyTwoOctXrayPeakSineGaussianNineteenNinetyFiveCircHzULNinety  & \ShortDurTwentyTwentyTwoOctXrayPeakRingdownFifteenNinetyaHzULNinety  & \ShortDurTwentyTwentyTwoOctXrayPeakRingdownFifteenNinetybHzULNinety  & \ShortDurTwentyTwentyTwoOctXrayPeakRingdownTwentyTwentyaHzULNinety  & \ShortDurTwentyTwentyTwoOctXrayPeakRingdownTwentyTwentybHzULNinety  \\
    2022 October 14            & Compact      & \ShortDurTwentyTwentyTwoOctFRBShortSineGaussianThirtyFiveSixtyHzULNinety  & \ShortDurTwentyTwentyTwoOctFRBShortSineGaussianSixteenHundredCircHzULNinety  & \ShortDurTwentyTwentyTwoOctFRBShortSineGaussianNineteenNinetyFiveCircHzULNinety  & \ShortDurTwentyTwentyTwoOctFRBShortRingdownFifteenNinetyaHzULNinety  & \ShortDurTwentyTwentyTwoOctFRBShortRingdownFifteenNinetybHzULNinety  & \ShortDurTwentyTwentyTwoOctFRBShortRingdownTwentyTwentyaHzULNinety  & \ShortDurTwentyTwentyTwoOctFRBShortRingdownTwentyTwentybHzULNinety  \\
    2022 October 14            & Extended   & \ShortDurTwentyTwentyTwoOctFRBLongSineGaussianThirtyFiveSixtyHzULNinety   & \ShortDurTwentyTwentyTwoOctFRBLongSineGaussianSixteenHundredCircHzULNinety   & \ShortDurTwentyTwentyTwoOctFRBLongSineGaussianNineteenNinetyFiveCircHzULNinety   & \ShortDurTwentyTwentyTwoOctFRBLongRingdownFifteenNinetyaHzULNinety   & \ShortDurTwentyTwentyTwoOctFRBLongRingdownFifteenNinetybHzULNinety   & \ShortDurTwentyTwentyTwoOctFRBLongRingdownTwentyTwentyaHzULNinety   & \ShortDurTwentyTwentyTwoOctFRBLongRingdownTwentyTwentybHzULNinety   \\
    2022 October 14 glitch 2   & Extended & \ShortDurTwentyTwentyTwoOctGlitchTwoSineGaussianThirtyFiveSixtyHzULNinety & \ShortDurTwentyTwentyTwoOctGlitchTwoSineGaussianSixteenHundredCircHzULNinety & \ShortDurTwentyTwentyTwoOctGlitchTwoSineGaussianNineteenNinetyFiveCircHzULNinety & \ShortDurTwentyTwentyTwoOctGlitchTwoRingdownFifteenNinetyaHzULNinety & \ShortDurTwentyTwentyTwoOctGlitchTwoRingdownFifteenNinetybHzULNinety & \ShortDurTwentyTwentyTwoOctGlitchTwoRingdownTwentyTwentyaHzULNinety & \ShortDurTwentyTwentyTwoOctGlitchTwoRingdownTwentyTwentybHzULNinety \\
    2022 December 01            & Extended   & \ShortDurTwentyTwentyTwoDecFRBLongSineGaussianThirtyFiveSixtyHzULNinety   & \ShortDurTwentyTwentyTwoDecFRBLongSineGaussianSixteenHundredCircHzULNinety   & \ShortDurTwentyTwentyTwoDecFRBLongSineGaussianNineteenNinetyFiveCircHzULNinety   & \ShortDurTwentyTwentyTwoDecFRBLongRingdownFifteenNinetyaHzULNinety   & \ShortDurTwentyTwentyTwoDecFRBLongRingdownFifteenNinetybHzULNinety   & \ShortDurTwentyTwentyTwoDecFRBLongRingdownTwentyTwentyaHzULNinety   & \ShortDurTwentyTwentyTwoDecFRBLongRingdownTwentyTwentybHzULNinety   \\
    \bottomrule
    \end{tabular}
\end{table*}

%% file: P2000488_v24_edited.tex
We thank Teruaki Enoto, Chin-Ping Hu, and George Younes for assistance with their X-ray results.
We also thank Vicky Kaspi, Kiyoshi Masui, and Kaitlyn Shin for helpful discussions, and the anonymous referee for their useful comments.

The gravitational-wave data analyzed in this paper is available on the
Gravitational Wave Open Science Center at \citet{gwosc_data}.
The scripts and data used to produce the figures in this paper are available at \citet{zenodo_release}.

This material is based upon work supported by NSF’s LIGO Laboratory, which is a
major facility fully funded by the National Science Foundation.
The authors also gratefully acknowledge the support of
the Science and Technology Facilities Council (STFC) of the
United Kingdom, the Max-Planck-Society (MPS), and the State of
Niedersachsen/Germany for support of the construction of Advanced LIGO
and construction and operation of the GEO\,600 detector.
Additional support for Advanced LIGO was provided by the Australian Research Council.
The authors gratefully acknowledge the Italian Istituto Nazionale di Fisica Nucleare (INFN),
the French Centre National de la Recherche Scientifique (CNRS) and
the Netherlands Organization for Scientific Research (NWO)
for the construction and operation of the Virgo detector
and the creation and support  of the EGO consortium.
The authors also gratefully acknowledge research support from these agencies as well as by
the Council of Scientific and Industrial Research of India,
the Department of Science and Technology, India,
the Science \& Engineering Research Board (SERB), India,
the Ministry of Human Resource Development, India,
the Spanish Agencia Estatal de Investigaci\'on (AEI),
the Spanish Ministerio de Ciencia, Innovaci\'on y Universidades,
the European Union NextGenerationEU/PRTR (PRTR-C17.I1),
the ICSC - CentroNazionale di Ricerca in High Performance Computing, Big Data
and Quantum Computing, funded by the European Union NextGenerationEU,
the Comunitat Auton\`oma de les Illes Balears through the Direcci\'o General de Recerca, Innovaci\'o i Transformaci\'o Digital with funds from the Tourist Stay Tax Law ITS 2017-006,
the Conselleria d'Economia, Hisenda i Innovaci\'o, the FEDER Operational Program 2021-2027 of the Balearic Islands,
the Conselleria d'Innovaci\'o, Universitats, Ci\`encia i Societat Digital de la Generalitat Valenciana and
the CERCA Programme Generalitat de Catalunya, Spain,
the Polish National Agency for Academic Exchange,
the National Science Centre of Poland and the European Union – European Regional
Development Fund;
the Foundation for Polish Science (FNP),
the Polish Ministry of Science and Higher Education,
the Swiss National Science Foundation (SNSF),
the Russian Science Foundation,
the European Commission,
the European Social Funds (ESF),
the European Regional Development Funds (ERDF),
the Royal Society,
the Scottish Funding Council,
the Scottish Universities Physics Alliance,
the Hungarian Scientific Research Fund (OTKA),
the French Lyon Institute of Origins (LIO),
the Belgian Fonds de la Recherche Scientifique (FRS-FNRS),
Actions de Recherche Concertées (ARC) and
Fonds Wetenschappelijk Onderzoek – Vlaanderen (FWO), Belgium,
the Paris \^{I}le-de-France Region,
the National Research, Development and Innovation Office of Hungary (NKFIH),
the National Research Foundation of Korea,
the Natural Science and Engineering Research Council of Canada (NSERC),
the Canadian Foundation for Innovation (CFI),
the Brazilian Ministry of Science, Technology, and Innovations,
the International Center for Theoretical Physics South American Institute for Fundamental Research (ICTP-SAIFR),
the Research Grants Council of Hong Kong,
the National Natural Science Foundation of China (NSFC),
the Israel Science Foundation (ISF),
the US-Israel Binational Science Fund (BSF),
the Leverhulme Trust,
the Research Corporation,
the National Science and Technology Council (NSTC), Taiwan,
the United States Department of Energy,
and
the Kavli Foundation.
The authors gratefully acknowledge the support of the NSF, STFC, INFN and CNRS for provision of computational resources.

This work was supported by MEXT,
the JSPS Leading-edge Research Infrastructure Program,
JSPS Grant-in-Aid for Specially Promoted Research 26000005,
JSPS Grant-in-Aid for Scientific Research on Innovative Areas 2905: JP17H06358,
JP17H06361 and JP17H06364,
JSPS Core-to-Core Program A,
Advanced Research Networks,
JSPS Grants-in-Aid for Scientific Research (S) 17H06133 and 20H05639,
JSPS Grant-in-Aid for Transformative Research Areas (A) 20A203: JP20H05854,
the joint research program of the Institute for Cosmic Ray Research,
the University of Tokyo,
the National Research Foundation (NRF),
the Computing Infrastructure Project of Global Science experimental Data hub
Center (GSDC) at KISTI,
the Korea Astronomy and Space Science Institute (KASI),
the Ministry of Science and ICT (MSIT) in Korea,
Academia Sinica (AS),
the AS Grid Center (ASGC) and the National Science and Technology Council (NSTC)
in Taiwan under grants including the Rising Star Program and Science Vanguard
Research Program,
the Advanced Technology Center (ATC) of NAOJ,
and the Mechanical Engineering Center of KEK.

Additional acknowledgements for support of individual authors may be found in the following document: \\
\texttt{https://dcc.ligo.org/LIGO-M2300033/public}.
For the purpose of open access, the authors have applied a Creative Commons Attribution (CC BY)
license to any Author Accepted Manuscript version arising.
We request that citations to this article use 'A. G. Abac {\it et al.} (LIGO-Virgo-KAGRA Collaboration), ...' or similar phrasing, depending on journal convention.